\documentclass[a4paper,10pt,twocolumn]{article}

\usepackage{graphicx}
\usepackage{cite}

\newcommand{\sect}[1]{\vspace*{\baselineskip}\noindent\textbf{#1}}
\newcommand{\subsect}[1]{\vspace*{0.5\baselineskip}\noindent\textbf{#1}}

\newcommand{\citeS}[1]{ \cite{#1}}
\newcommand{\citeR}[1]{ \cite{#1}}

\newcommand{\VT}[2]{\left[{\begin{array}{c}#1\\#2\end{array}}\right]}
\newcommand{\MT}[4]{\left[\begin{array}{ccc}#1&\:&#2\\#3&\:&#4\end{array}\right]}

\begin{document}
\twocolumn[{\center
	{\bf {\Large
Experimental demonstration of spinor slow light
	}}

	\parbox{14.5cm}{{\center
Meng-Jung Lee,$^1$ Julius Ruseckas,$^2$ Chin-Yuan Lee,$^1$ Via\v{c}eslav Kudria\v{s}ov,$^2$ Kao-Fang Chang,$^1$ Hung-Wen Cho,$^1$ Gediminas Juzeli\={u}nas,$^{2}$ and Ite A. Yu$^{1,\ast}$ \\
		}}
	\vspace*{-0.75\baselineskip}

	\parbox{15cm}{\center{\it{\small
$^1$Department of Physics and Frontier Research Center on Fundamental and Applied Sciences of Matters, \\
	National Tsing Hua University, Hsinchu 30013, Taiwan \\
$^2$Institute of Theoretical Physics and Astronomy, \\
	Vilnius University, A. Go\v{s}tauto 12, Vilnius 01108, Lithuania \\
	}}}
	\vspace*{\baselineskip}

	\parbox{15cm}{\hspace*{0.15cm}
Slow light based on the effect of electromagnetically induced transparency is of great interest due to its applications in low-light-level nonlinear optics and quantum information manipulation. The previous experiments all dealt with the single-component slow light. Here we report the experimental demonstration of two-component or spinor slow light using a double tripod atom-light coupling scheme. The scheme involves three atomic ground states coupled to two excited states by six light fields. The oscillation due to the interaction between the two components was observed. Based on the stored light, our data showed that the double tripod scheme behaves like the two outcomes of an interferometer enabling precision measurements of frequency detuning. We experimentally demonstrated a possible application of the double tripod scheme as quantum memory/rotator for the two-color qubit. Our study also suggests that the spinor slow light is a better method than a widely-used scheme in the nonlinear frequency conversion.
	}
	\vspace*{1.5\baselineskip}
}]

\footnotetext{\\ $^{\ast}$yu@phys.nthu.edu.tw}

\newcommand{\FigOne}{
	\begin{figure}[!t] 
	\setlength{\leftmargini}{0cm}
	{\center\includegraphics[width=8.5cm]{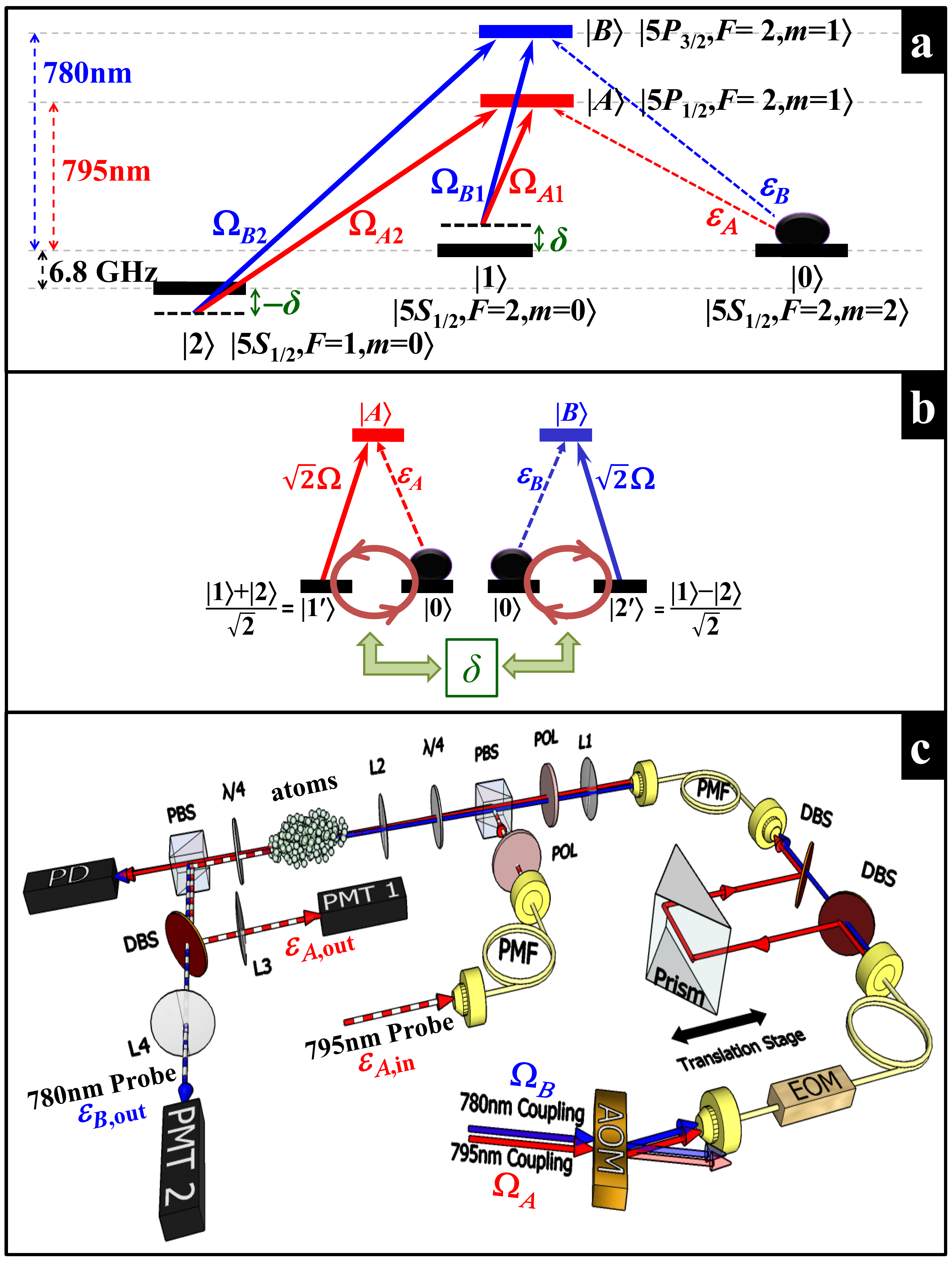}\\}
	\textbf{Figure 1 $\vert$ 
Transition diagram and experimental setup.}
	\textbf{a,}
Relevant energy levels and laser excitations in the double-tripod (DT) system for $^{87}$Rb atoms. Here $\varepsilon_A$ and $\varepsilon_B$ represent the probe fields; $\Omega_{A1}$, $\Omega_{A2}$, $\Omega_{B1}$ and $\Omega_{B2}$ indicate the coupling fields; $\delta$ (or $-\delta$) is the two-photon detuning with respect to the Raman transition between $|0\rangle$ and $|1\rangle$ (or $|2\rangle$).
	\textbf{b,} Two coupled $\Lambda$ systems with an effective Rabi frequency $\sqrt{2}\Omega$. The schema is equivalent to the DT system with Rabi frequencies of the four coupling fields having the same amplitude $\Omega$ and relative phase $\theta=\pi$. The two-photon detuning $\delta$ of the DT system introduces coupling between the effective $\Lambda$ systems.
	\textbf{c,}
Schematic experimental setup. AOM: acousto-optic modulator; EOM: electro-optic modulator; DBS: dichroic beam splitter; PBS: polarizing beam splitter; PMF: polarization-maintained optical fiber; POL: polarizer; L1-L4: lenses with focal lengths of 300, 200, 300 and 500 mm, respectively; $\lambda$/4: quarter-wave plate; PMT: photo multiplier; PD: photo detector.
	\end{figure}
}
\newcommand{\FigTwo}{
	\begin{figure}[!t] 
	\setlength{\leftmargini}{0cm}
	{\center\includegraphics[width=7.5cm]{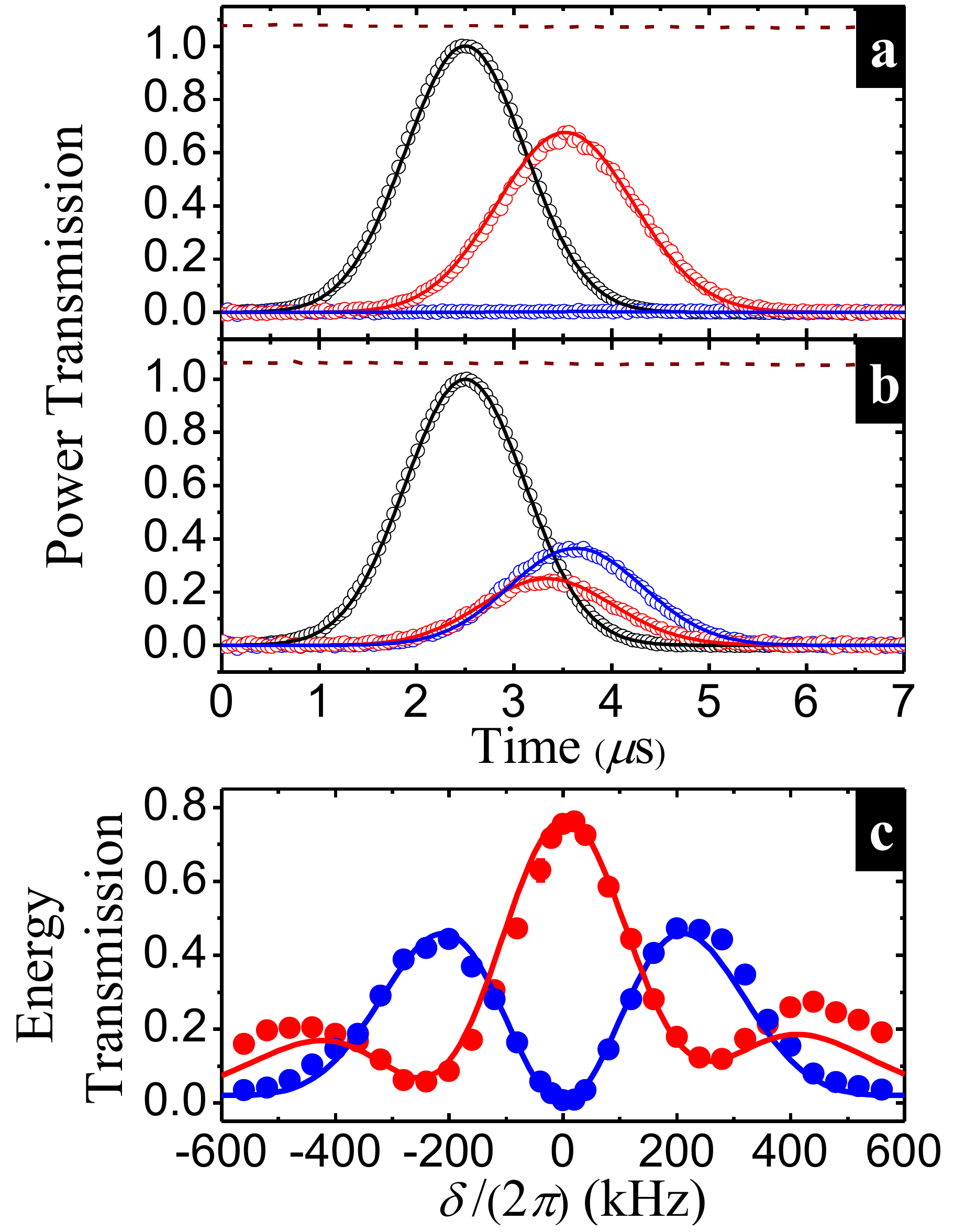}\\}
	\textbf{Figure 2 $\vert$ 
Oscillation phenomenon in the double-tripod scheme.}
	\textbf{a,b,}
The output powers of $\varepsilon_A$ and $\varepsilon_B$ versus time for $\varepsilon_A$ being the only input, where $\delta = 0$ in \textbf{a} and 2$\pi$$\times$160 kHz in \textbf{b}. Black and red circles are experimental data of the input and output $\varepsilon_A$; blue circles are those of the output $\varepsilon_B$. Dashed lines represent data of the four coupling fields. Solid lines are theoretical predictions.
	\textbf{c,}
Energy transmissions of the two probes as functions of $\delta$. Circles are experimental data and solid lines are theoretical predictions. In the theoretical calculation, $\alpha$ = 20, $\Omega_{A1}$ = $\Omega_{A2}$ = $\Omega_{B1}$ = $\Omega_{B2}$ = 0.51$\Gamma$, $\gamma_1$ = 0, $\gamma_2$= $3.7\times10^{-3}$$\Gamma$ and $\Delta_k L$ = 0.6, where $\Gamma$ = $2\pi$$\times$6 MHz. These parameters were determined by two ordinary (single-$\Lambda$) EIT and two double-$\Lambda$ measurements. Error bars are about the size of data points.
	\end{figure}
}
\newcommand{\FigThree}{
	\begin{figure}[!t] 
	\setlength{\leftmargini}{0cm}
	{\center\includegraphics[width=7.5cm]{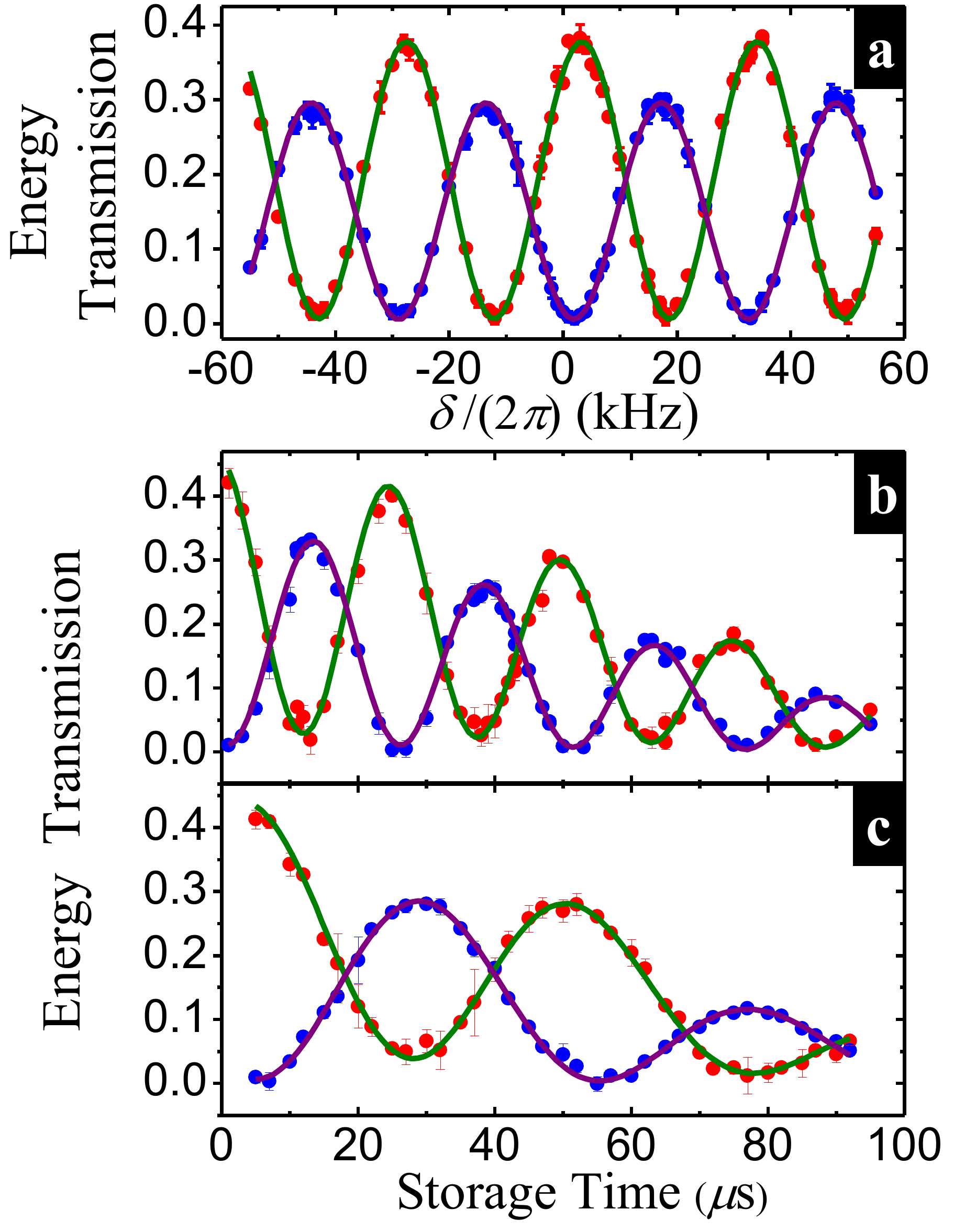}\\}
	\textbf{Figure 3 $\vert$ 
Light storage in the double-tripod scheme with the probe $\varepsilon_A$ being the only input.}
	\textbf{a,}
After a storage time of 15 $\mu$s, retrieved energies of $\varepsilon_A$ (red circles) and $\varepsilon_B$ (blue circles) as functions of $\delta$. Green and purple lines are the best fits, determining the oscillation period equal to 2$\pi$$\times$(30.8$\pm$0.1) kHz.
	\textbf{b,c,}
Retrieved energies as functions of the storage time at two values of $\delta$ differing by 2$\pi$$\times$10.0 kHz. The best fits in \textbf{b} (or \textbf{c}) determine the oscillation period equal to 49.9$\pm$0.3 (or 25.3$\pm$0.1) $\mu$s and the decay time constant equal to 76.8$\pm$1.2 (or 75.8$\pm$1.5) $\mu$s.
	\end{figure}
}
\newcommand{\FigFour}{
	\begin{figure}[!t] 
	\setlength{\leftmargini}{0cm}
	{\center\includegraphics[width=7.5cm]{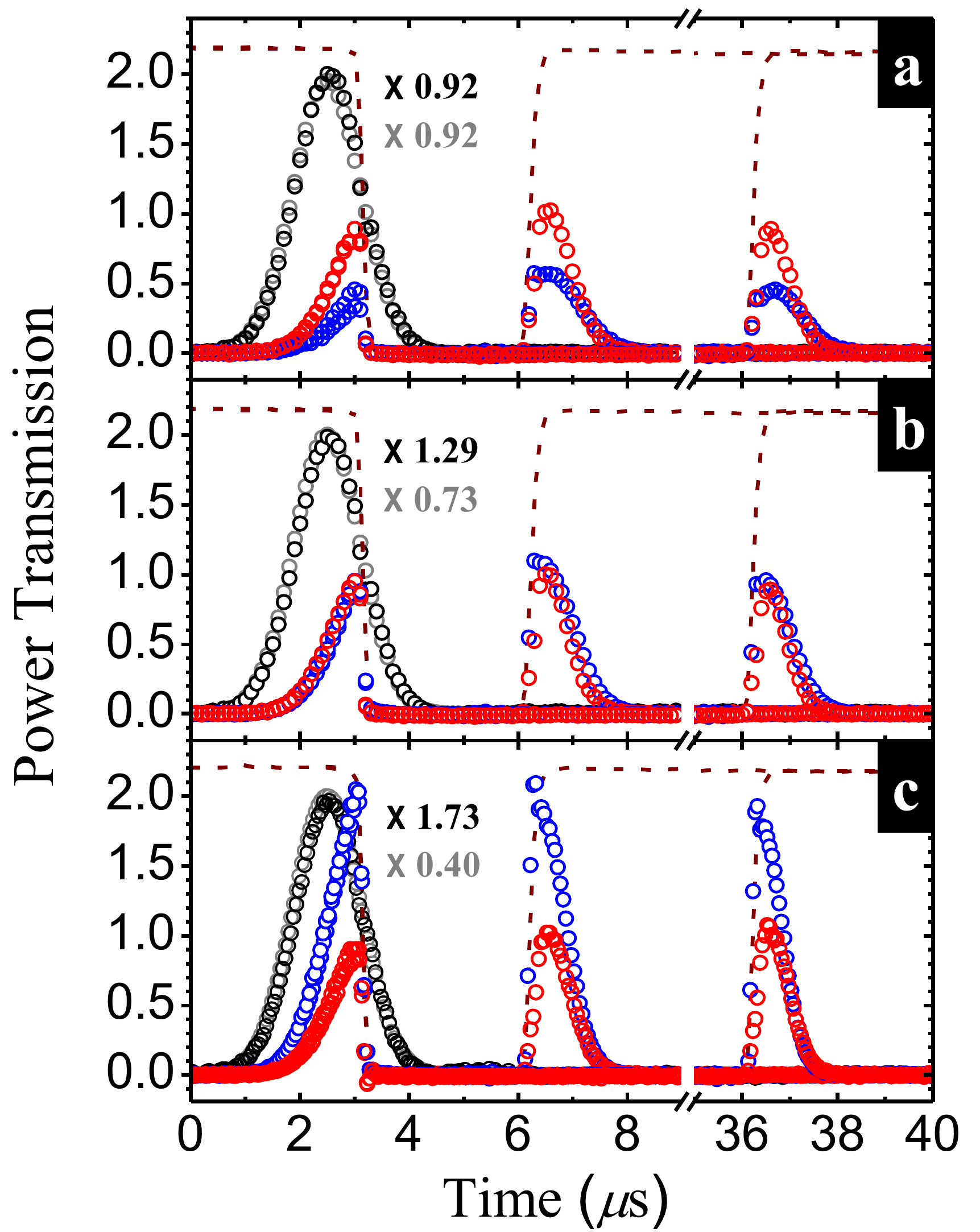}\\}
	\textbf{Figure 4 $\vert$ 
Storage and retrieval of two-color light pulses in the double-tripod scheme.}
The detuning $\delta$ was set to zero during the storage time. Black and gray circles are the input $\varepsilon_A$ and $\varepsilon_B$ pulses scaled up or down by the factors shown in the plots; red and blue circles are the two retrieved pulses after storage times of 3 and 33 $\mu$s. In \textbf{a}, \textbf{b} and \textbf{c}, energy ratios of the two retrieved pulses after the storage time of 3 (33) $\mu$s are 1.5 (1.5), 0.84 (0.92) and 0.55 (0.52), respectively.
	\end{figure}
}
\newcommand{\FigFive}{	
	\begin{figure}[!t] 
	\setlength{\leftmargini}{0cm}
	{\center\includegraphics[width=8.5cm]{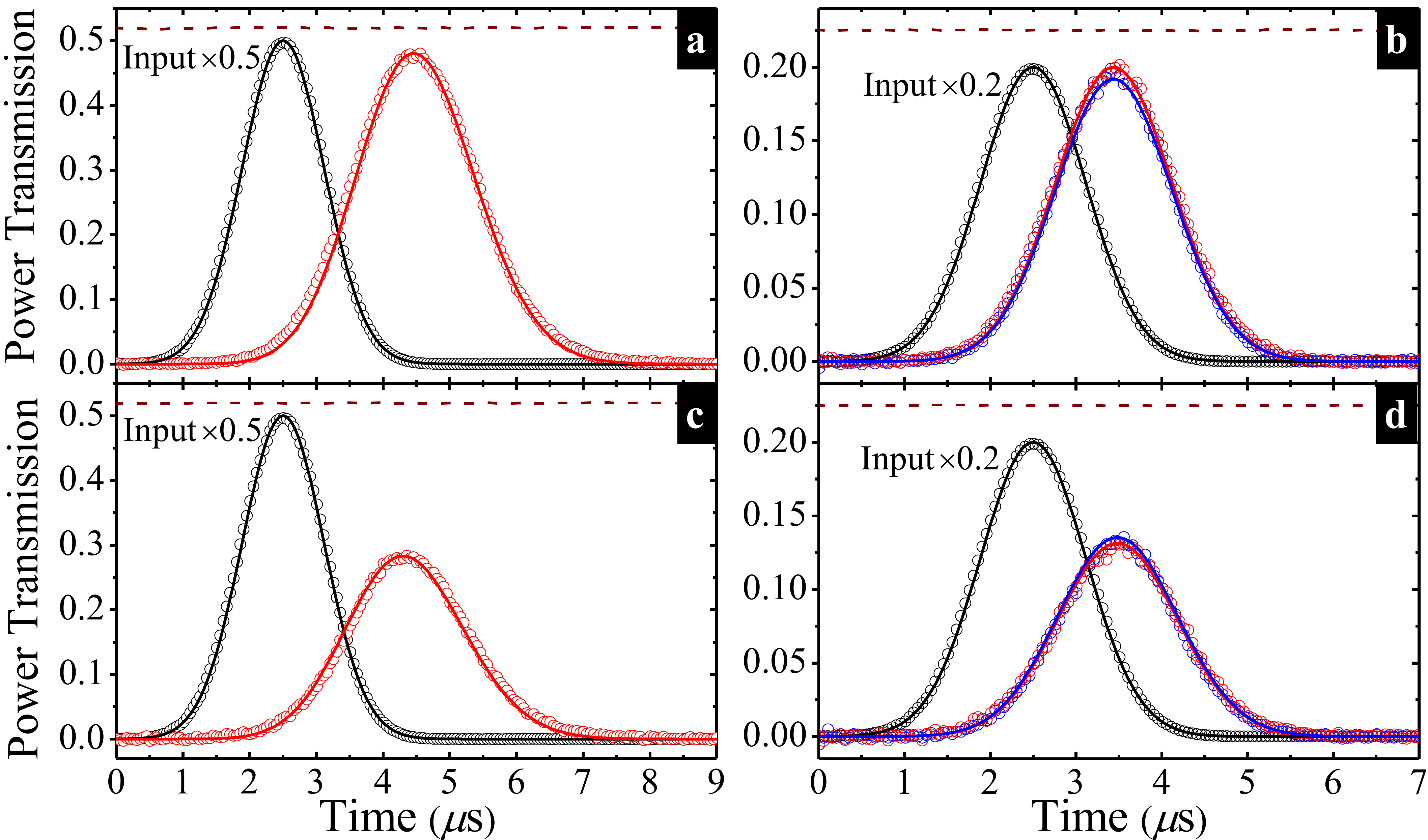}\\}
	\textbf{Figure 5 $\vert$ 
Determination of experimental parameters.
	}
	\textbf{a,(or c,)}
The slow light output of $\varepsilon_A$ under the coupling field of $\Omega_{A1}$ (or $\Omega_{A2}$) in the ordinary (single-$\Lambda$) EIT scheme.
	\textbf{b,(or d,)}
The slow light outputs of $\varepsilon_A$ and $\varepsilon_B$ under the coupling fields of $\Omega_{A1}$ and $\Omega_{B1}$ (or $\Omega_{A2}$ and $\Omega_{B2}$) in the double-$\Lambda$ scheme with $\varepsilon_A$ being the only input. Black and red circles are the experimental data of the input and output of $\varepsilon_A$; blue circles are those of the output of $\varepsilon_B$; dashed lines represent the data of the coupling field timing; solid lines are the theoretical predictions. In the theoretical calculation, $\theta$ = $\pi$, $\alpha$ = 20$\pm$1, $|\Omega_{A1}|$ = $|\Omega_{A2}|$ = (0.51$\pm$0.02)$\Gamma$, $|\Omega_{B1}|$ =  $|\Omega_{B2}|$ = (0.52$\pm$0.015)$\Gamma$, $\gamma_1 \leq$ 3$\times$10$^{-4}$$\Gamma$ or is effectively 0, $\gamma_2$ = 3.7$\times$10$^{-3}$$\Gamma$ and $\Delta_k L$ = 0.6.
	\end{figure}
}

Over the last decade there have been significant advances in studying the slow\citeS{SL1,SL2,SL3,SL4,SL5,SL6,SL7,SL8,SL9,SL10,SL11,SL12}, stored\citeS{LS1,LS2,LS3,LS4,LS5,LS6,LS7,OurPRL13,LS60s} and stationary\citeS{SLP1,SLP2,OurPRL09,SLP4} light stimulated by applications to low-light-level nonlinear optics\citeS{XPM1,XPM2,XPM3,AOS1,AOS2,rEIT1,rEIT2,rEIT3,YccPRL,OurPRL12} and quantum information manipulation\citeS{DLCZ,BPG,QM1,QM2,QM3,QM4,QM5}. The slow and stationary light, forming due to the electromagnetically induced transparency (EIT) effect\citeS{EIT1,EIT2,EIT3}, greatly enhance the light-matter interaction and enable nonlinear optical processes to achieve significant efficiency even at single-photon level\citeS{XPM1,XPM2,XPM3,AOS1,AOS2,rEIT1,rEIT2,rEIT3,YccPRL,OurPRL12}. The storage of light using the dynamic EIT scheme transfers quantum states between photons and atoms, serving as quantum memory for photons\citeS{QM1,QM2,QM3,QM4,QM5}. The EIT-related research has made a great impact on the nonlinear optics and quantum information science\citeS{XPM1,XPM2,XPM3,AOS1,AOS2,rEIT1,rEIT2,rEIT3,YccPRL,OurPRL12,DLCZ,BPG,QM1,QM2,QM3,QM4,QM5}.

The two-component or spinor slow light (SSL) using a double tripod (DT) atom-light coupling scheme\citeS{DT-Dirac,DT-PBG,DT-OAM,DT-SLP} exhibits a number of additional distinct features. The SSL can lead to interesting phenomena, such as formation of the quasi-particles exhibiting Dirac spectra\citeS{DT-Dirac,DT-PBG} or oscillations between the two components\citeS{DT-PBG,DT-OAM}. It can also be exploited in designing novel photonic devices, e.g. quantum memory/rotator for two-color qubits, interferometers for sensitive measurements and high-efficiency media for nonlinear frequency conversion, as it will be discussed later in the article. Here we report the first experimental demonstration of the SSL.

Our experimental study of the SSL makes use of the DT transition scheme and was carried out with laser-cooled $^{87}$Rb atoms. Details of the experimental setup can be found in the Methods section. The DT level scheme consists of three atomic ground states of $|0\rangle$, $|1\rangle$ and $|2\rangle$ and two excited states of $|A\rangle$ and $|B\rangle$, as depicted in Fig.~1a. One probe field (with the Rabi frequency $\varepsilon_A$) and two coupling fields (with the Rabi frequencies $\Omega_{A1}$ and $\Omega_{A2}$) drive the transitions from $|0\rangle$, $|1\rangle$ and $|2\rangle$ to $|A\rangle$, respectively, to form the first tripod configuration. Another probe field ($\varepsilon_B$) and the other two coupling fields ($\Omega_{B1}$ and $\Omega_{B2}$) drive the transitions from the same ground states to $|B\rangle$ to form the second tripod configuration. The DT scheme is a combination of two single-tripod schemes\citeS{ST1,ST2,ST3,ST4,ST5}, but its physics is more abundant due to the interaction between the two components of light coupled with two atomic coherences\citeS{DT-Dirac,DT-PBG,DT-OAM,DT-SLP}.

In this work, we observe the oscillation between the two slow light components controlled by the two-photon detuning. In a proof-of-principle measurement, our data show that the DT scheme for the light storage behaves like the two outcomes of an interferometer enabling measurements of the frequency detuning with the precision on the order of 100 Hz. Finally, we experimentally demonstrate a possible application of the DT scheme as quantum memory/rotator for the two-color qubit, i.e. the superposition state of two frequency modes. 

\FigOne

\sect{Results}

\subsect{Theoretical background.} 
In the DT system\citeS{DT-Dirac,DT-PBG,DT-OAM,DT-SLP}, the dynamics of the two probe fields and two atomic coherences can be described by the Maxwell-Bloch equations\citeS{DT-Dirac,DT-SLP}: 
\begin{equation}
	\frac{1}{c}\frac{\partial}{\partial t} \VT{\varepsilon_A}{\varepsilon_B}
		+\frac{\partial}{\partial z} \VT{\varepsilon_A}{\varepsilon_B} 
		=i\frac{\alpha\Gamma}{2L} \VT{\rho_{A}}{\rho_{B}},
\end{equation}
\begin{equation}
	\frac{\partial}{\partial t} \VT{\rho_{A}}{\rho_{B}} 
		=\frac{i}{2} \VT{\varepsilon_A}{\varepsilon_B} 
		+\frac{i}{2} \MT{\Omega_{A1}}{\Omega_{A2}}{\Omega_{B1}}{\Omega_{B2}}
		\VT{\rho_{1}}{\rho_{2}}
		-\frac{\Gamma}{2} \VT{\rho_{A}}{\rho_{B}},
\end{equation}
\begin{equation}
	\frac{\partial}{\partial t} \VT{\rho_{1}}{\rho_{2}} 
		=\frac{i}{2} 
		\MT{\Omega_{A1}^*}{\Omega_{B1}^*}{\Omega_{A2}^*}{\Omega_{B2}^*} 
		\VT{\rho_{A}}{\rho_{B}} 
		+\MT{i\delta}{0}{0}{-i\delta} \VT{\rho_{1}}{\rho_{2}},
\end{equation}
where $\rho_{A}$ (or $\rho_{B}$) is the optical coherence corresponding to the probe transition of $|0\rangle \rightarrow |A\rangle$ (or $|0\rangle \rightarrow |B\rangle$), $\rho_{1}$ (or $\rho_{2}$) is the ground-state coherence between $|0\rangle$ and $|1\rangle$ (or $|2\rangle$), $\Gamma$ is the spontaneous decay rate of the excited states, $\alpha$ is the optical density (OD) of the medium with the length $L$, and $\delta$ is the two-photon detuning, as illustrated in Fig.~1a. To reach the above equations, the probe fields are assumed to be much weaker than the coupling ones. In that case most atomic population is in the ground state $|0\rangle$, and one can treat the probe fields as a perturbation. All fast-oscillation exponential factors associating with center frequencies and wave vectors have been eliminated from the equations, and only slowly-varying amplitudes are retained.

To simplify the discussion, let the complex Rabi frequencies of the four coupling fields $\Omega_n = \Omega e^{i\theta_n}$ have the same constant amplitude of $\Omega$ and various phases $\theta_n$, where $n = A1$, $A2$, $B1$ or $B2$. We define $\theta \equiv (\theta_{A1}-\theta_{A2})-(\theta _{B1}-\theta _{B2})$ to be a relative phase among the four coupling fields. By changing $\theta$ one can substantially alter the dispersions and other properties of the two-component slow-light modes\citeS{DT-Dirac,DT-PBG,DT-OAM}. We will focus on the case where $\theta = \pi$. According to Eqs.~(1)-(3), the relation between the input and output probe fields of the continuous wave is then given by (see Supplementary Note 1)
\begin{equation}
	\VT{\varepsilon_{A,{\rm out}}}{\varepsilon_{B,{\rm out}}} 
		= {\rm e}^{-2\phi^2/\alpha}
		\MT{\cos\phi}{\:-\sin\phi}{\sin\phi}{\:\cos\phi}
		\VT{\varepsilon_{A,{\rm in}}}{\varepsilon_{B,{\rm in}}},
\end{equation}
with 
\begin{equation}
	\phi = \frac{\alpha\Gamma}{2\Omega^2} \delta .
	\label{eq:phiSL}
\end{equation}
As only one probe field is present in the input, e.g. $\varepsilon_{A,{\rm in}} = 1$ and $\varepsilon_{B,{\rm in}} = 0$, the probe transmissions are
\begin{equation}
	\VT{|\varepsilon_{A,{\rm out}}|^2}{|\varepsilon_{B,{\rm out}}|^2} 
		= {\rm e}^{-4\phi^2/\alpha} \VT{\cos^2\phi}{\sin^2\phi}.
	\label{eq:DTosc}
\end{equation}
Thus oscillations between the two modes show up at the output of the medium.

The oscillation phenomenon as well as the interaction of the probe fields with the atomic medium is determined by the four coupling fields. Figure~1b illustrates a physical picture of the role played by the coupling fields for the case where $\theta$ = $\pi$. The coupling fields couple the excited state $|A\rangle$ ($|B\rangle$) to a symmetric (anti-symmetric) superposition of the atomic ground states $|1\rangle$ and $|2\rangle$ labeled by $|1'\rangle$ ($|2'\rangle$). Supplementary Note 2 explains in details how the DT system is equivalent to the two coupled $\Lambda$ systems. The detuning $\delta$ introduces the coupling between the newly defined ground states $|1'\rangle$ and $|2'\rangle$, as one can see in Fig.~1b. This leads to the interaction between the probe fields and hence to their oscillations during the propagation in the atomic medium. 

The SSL oscillations can be also observed for the relative phase $\theta$ of coupling fields other than $\pi$. However, at $\theta$ = 0 the excited states$|A\rangle$ and $|B\rangle$ are coupled to the same (symmetric) superposition of $|1\rangle$ and $|2\rangle$  (see Supplementary Note 3). In that case the DT system becomes equivalent to the double-$\Lambda$ system, and no oscillation can occur between the probe fields. Larger $\theta$ makes the oscillations more prominent, and $\theta$ = $\pi$ gives the maximum contrast or difference between two output probe fields at a small $\delta$ (see Supplementary Note 4). For this reason, $\theta$ = $\pi$ was chosen in the experiment.

According to Eqs.~(5) and (6), the detuning $\delta$ not only causes the probe fields to oscillate, but also reduces the total output energy. In the EIT spectrum, the two-photon resonance (i.e., a condition where the frequency difference between the probe and coupling fields is equal to that between two ground states driven by them) corresponds to the transparency peak or the maximum transmission. A larger two-photon detuning makes less transmission. In the present EIT system, the two-photon detuning and coupling Rabi frequency correspond to $\delta$ and $\sqrt{2}\Omega$. Based on Eq.~(27) in Ref.~\citeR{EIT3}, the transmission around the EIT peak is given by $\exp[-4\alpha\delta^2\Gamma^2/(\sqrt{2}\Omega)^4]$ which is exactly the exponential decay term in Eq.~(6).

\subsect{Nonlinear frequency conversion.}
By varying one of the three factors, the two-photon detuning $\delta$, optical density (OD) $\alpha$ and coupling Rabi frequency $\Omega$, while keeping the other two fixed, we can make the two slow-light outputs oscillate alternatively. The oscillations are accompanied by the decay of the total output energy, since the EIT is degraded due to the detuning $\delta$. Nevertheless, a large OD can significantly reduce the decay over an oscillation period. Regarding the nonlinear optical process that converts light from one frequency to another, one can employ the DT scheme at $\phi = \pi/2$. According to Eq.~(\ref{eq:DTosc}), an OD of 250 enables the conversion efficiency of about 96\%. The same efficiency requires an OD of 500 in the widely-used double-$\Lambda$ scheme\citeS{FWM-YFC14}. Hence, the DT scheme is a new and advantageous method of nonlinear frequency conversion.

\FigTwo

\subsect{Oscillation of spinor slow light.} 
We first demonstrated the oscillation phenomenon resulting from the SSL in the DT scheme. As only one probe pulse ($\varepsilon_A$) was sent to the input and the four coupling fields were constantly present, the outputs of both probe pulses were measured at different two-photon detunings. Figures~2a and 2b show the two output powers as functions of time at $\delta = 0$ and  $\delta$ = 2$\pi$$\times$160 kHz. At $\delta = 0$, $\varepsilon_B$ will be generated by $\varepsilon_A$ if $\theta \neq \pi$ and, $\varepsilon_B$ will not be generated if $\theta = \pi$ (see Supplementary Note 5). Because of this condition, we were able to properly set $\theta = \pi$ by minimizing $\varepsilon_B$'s output. The energy transmissions of $\varepsilon_A$ and $\varepsilon_B$ as functions of $\delta$ are shown in Fig.~2c. Two outputs oscillate alternatively; when one reaches minima the other becomes maxima and vice versa. Their total transmitted energy decays as $|\delta|$ increases, because the detuning away from the EIT resonance is associated with the losses. The oscillation phenomenon behaves qualitatively as described by Eqs.~(\ref{eq:phiSL}) and (\ref{eq:DTosc}). 

Quantitatively, we compare the data with the predictions by numerically solving Eqs.~(1)-(3)  as shown in Figs.~2a-2c. To better describe the experimental condition, we also considered the phase mismatch $\Delta_k$ and the dephasing rates $\gamma_1$ and $\gamma_2$ of the ground-state coherences $\rho_1$ and $\rho_2$ in the calculation. The following two terms
\begin{displaymath}
	\MT{i\Delta_k/2}{0}{0}{-i\Delta_k/2} \VT{\varepsilon_A}{\varepsilon_B}
	\; {\rm and} \;
	\MT{-\gamma_1}{0}{0}{-\gamma_2} \VT{\rho_{1}}{\rho_{2}}
\end{displaymath}
have been added to the left-hand side of Eq.~(1) and the right-hand side of Eq.~(3), respectively. In the first term, $\Delta_k \equiv (-{\bf\vec{k}}_{pA} +{\bf\vec{k}}_{cA} -{\bf\vec{k}}_{cB} +{\bf\vec{k}}_{pB}) \cdot {\bf\hat{z}}$ describes the effect of phase mismatch\citeS{PM}, where ${\bf\vec{k}}_{pA}$ and ${\bf\vec{k}}_{pB}$ are the wave vectors for the probe fields $\varepsilon_A$ and $\varepsilon_B$, and ${\bf\vec{k}}_{cA}$ and ${\bf\vec{k}}_{cB}$ are those for the coupling fields $\Omega_{A1}$ and $\Omega_{B1}$ (or $\Omega_{A2}$ and $\Omega_{B2}$). There are no free parameters in the calculation, in which all parameters were determined by two ordinary (single-$\Lambda$) EIT and two double-$\Lambda$ measurements (see the Methods section). In Fig.~2c, $\varepsilon_A$'s minima are not completely zero, and the probe transmission is slightly asymmetric for the positive and negative detunings. The nonzero minima are caused by $\Delta_k \neq 0$ and the finite frequency bandwidth of the input probe pulse, whereas the asymmetry results from the combination of $\Delta_k \neq 0$ and $\gamma_1 \neq \gamma_2$ (see Supplementary Note 6 for more details). For $|\delta| <$ 300 kHz, discrepancies between the data and predictions are comparable to the measurement uncertainty which is about the size of the data point. At a large detuning, some data point can deviate from the theoretical line significantly more than the uncertainty. The deviation may be due to the one-photon detunings of the 795 nm and 780 nm transitions, which cannot be determined accurately and has not been taken into account in the calculation. 

\subsect{Spinor-slow-light interferometer.}
The number of oscillation cycles can be considerably increased with the storage and retrieval of SSL. The idea is based on the intuition that the propagation time of the light pulses in the medium is equivalent to the storage time of motionless ones transformed into the atomic coherences $\rho_1$ and $\rho_2$. In Eq.~(\ref{eq:phiSL}) the quantity $t_d \equiv \alpha\Gamma/(2\Omega^2)$, representing the SSL propagation delay time, determines the mixing angle $\phi$ (= $t_d \delta$) between the two slow-light components. If the slow light is stored for a time $t_s \gg t_d$, the propagation time $t_d$ is to be replaced by the storage time $t_s$ in the phase of the SSL oscillation
\begin{equation}
	\phi = t_s \delta
	\label{eq:phiLS}
\end{equation}
without introducing extra losses. In this work, the propagation delay time $t_d$ was merely around 1 $\mu$s. With the storage time $t_s$ considerably longer than 1 $\mu$s, a small two-photon detuning (compared with the EIT bandwidth) can still result in a large oscillation phase. Figure~3a clearly demonstrates the above idea and verifies Eq.~(\ref{eq:phiLS}) at $t_s =$ 15 $\mu$s. The data of retrieved energies versus $\delta$ exhibit more than three oscillation cycles and show no oscillation amplitude attenuation. Both $\varepsilon_A$'s and $\varepsilon_B$'s minima with values of $0.017\pm 0.012$ are close to zero. The data of positive and negative values of $\delta$ are very symmetric except a small shift of the origin in $\delta$. This shift is due to the AC Stark effect induced by the coupling fields not present during the light storage. The data in Fig.~3a behave just like $\cos^2\phi$ and $\sin^2\phi$, where $\phi = (t_s+t_d)\delta +\phi_0$. Their best fits determine that the oscillation period is 2$\pi$$\times$(30.8$\pm$0.1) kHz. This period indicates $t_s + t_d \approx$ 16 $\mu$s which is in agreement with the actual value in the measurement.

\FigThree

The data of two output probe energies in Fig.~3a are similar to the two outcomes of an interferometer. According to Eq.~(\ref{eq:phiLS}), one can precisely determine the two-photon detuning $\delta$ with such DT interferometer. The decay time constant of stored light in this experiment was about 76 $\mu$s. Utilizing this sufficiently long time constant, we demonstrated a proof-of-principle measurement in Figs.~3b and 3c. Data in each figure were taken under a fixed $\delta$. We set the difference in the $\delta$ to be 2$\pi$$\times$10.0 kHz for the two figures. Note that the difference of the two values can be set much more accurately than their absolute values. We measured the retrieved energies against the storage time as shown in the figures. The best fits of the data determine the oscillation periods ($T_s$) of 49.9$\pm$0.3 and 25.3$\pm$0.1 $\mu$s. Based on Eq.~(\ref{eq:phiLS}), we have $\delta = \pi/T_s$ which gives the difference of the two $\delta$s equal to 2$\pi$$\times$9.7 kHz. The measured value of the difference is consistent with the actual one, showing that the light-storage DT scheme can be used to determine the detuning $\delta$ or anything that can affect $\delta$, such as light shifts, Zeeman shifts, etc. The precision demonstrated here is on the order of 100 Hz.

\FigFour

\subsect{Two-color qubits.}
The single-photon SSL can be considered as the qubit with the superposition state of two frequency modes or, simply, as the two-color qubit. Its wave function is given by
\begin{equation}
	|\psi\rangle  = a |1_{\omega_A},0_{\omega_B}\rangle
		+b |0_{\omega_A},1_{\omega_B}\rangle,
\end{equation}
where $a$ and $b$ are the probability amplitudes with the constraint of $|a|^2$ + $|b|^2$ = 1. Such a qubit can be produced by sending a single photon to the DT system. On the other hand, the DT system can be used as quantum memory/rotator for the qubits. In the classical limit, the above wave function corresponds to $a${\boldmath$\vec{\varepsilon}$}$_A$ $+$ $b${\boldmath$\vec{\varepsilon}$}$_B$, where {\boldmath$\vec{\varepsilon}$}$_A$ and {\boldmath$\vec{\varepsilon}$}$_B$ represent the basis or electric fields of the two frequency modes. Considering $(a,b) = (\cos\phi,\sin\phi)$, the generation of this wave function has been demonstrated in Fig.~3a. 

To see whether the light storage in the DT system can preserve $(a,b)$, we simultaneously sent both $\varepsilon_A$ and $\varepsilon_B$ pulses to the medium and made $\delta = 0$ during the storage. Figures~4a, 4b and 4c show the temporal profiles of the two retrieved pulses at different ratios of $a$ to $b$. In each figure, the shapes of the two retrieved pulses and their energy ratio after the storage time of 3 $\mu$s are very close to those after the storage time of 33 $\mu$s. Furthermore, one should be able to utilize a two-photon detuning applied during the storage to change $\phi$ or the $a$-to-$b$ ratio. The detuning is essentially the Zeeman or AC Stark shift induced by a pulse of magnetic field or by a far detuned microwave or laser pulse. The amount of the change is controlled by the product of the detuning and pulse duration as shown by the data in Fig.~3. Although our demonstrations were done with classical light, the results suggest that the light-storage DT scheme can be used as the quantum memory/rotator for two-color qubits.

\sect{Discussion} 

\vspace*{0.5\baselineskip} \noindent
We observed the two-component slow light or SSL in the DT system as demonstrated by the oscillation of the two output light pulses. The oscillations are induced by the coupling between the two ground-state coherences due to the non-zero detuning $\delta$, as shown in Fig.~1b. The data of the light-storage DT scheme were used to determine the two-photon detuning (or anything that can affect the detuning such as light shifts, Zeeman shifts, etc.) in the system with the satisfactory accuracy and precision. The sensitivity of our method is based on the fact that the light is stored in a superposition of two atomic levels, i.e. in two atomic coherences. Therefore, for sufficiently long storage times even a slight energy mismatch between these levels can lead to a large accumulated phase which can be detected by measuring the conversion of the regenerated light into another component. The storage time and the measurement precision in this work were on the orders of 100 $\mu$s and 100 Hz, respectively. In Ref.~\citeR{LS7} an optical dipole trap was employed to confine cold atoms, leading to the light-storage time of about 1~s. In principle, a longer storage time can result in a better precision of the measured frequency.

The two-color qubit is a superposition state of two frequency modes. We demonstrated a possible application of employing the DT scheme as quantum memory/rotator for the two-color qubits by utilizing a two-photon detuning during the storage time. Furthermore, the SSL may lead to interesting physics such as spinor Bose-Einstein condensation of dark-state polaritons\citeS{SLP2} and Dirac particles\citeS{DT-Dirac,DT-PBG}. It can also be used to achieve high conversion efficiencies in the nonlinear frequency conversion. This work is the first experiment of the SSL research, which may result in novel applications in quantum information manipulation, precision measurement and nonlinear optics.

Finally, we note that the current DT setup is arranged in the co-propagation configuration in which the four coupling fields propagate in the same direction and completely overlap. The phase match condition can be closely achieved in the co-propagation configuration. One can also use the counter-propagation configuration to form the SSL\citeS{DT-Dirac,DT-PBG}. In such a situation the coupling fields come from two opposite directions to interact with atoms. Yet a phase mismatch can be significant even at an optimum alignment of laser beams, which has been already seen in our previous experiments utilizing the counter-propagating configuration for other studies\citeS{OurPRL13,OurPRL12}. This may represent an obstacle in the SSL formation for the counter-propagating configuration. Regarding nonlinear frequency conversion, the co-propagation scheme\citeS{DT-OAM} allows for a higher efficiency as compared to the counter-propagating one\citeS{DT-PBG}.

\sect{Methods}

\subsect{Experimental setup.}
A cigar-shaped cloud of cold $^{87}$Rb atoms with the dimension of approximately 9$\times$2$\times$2 mm$^3$ was produced by a magneto-optical trap (MOT)\citeS{CigarMOT}. Typically, we can trap $10^9$ atoms with a temperature of about 300 $\mu$K in the MOT. Figure~1a shows the relevant energy levels and laser excitations in the experiment. The system of cold atoms is not a necessary condition for the SSL formation. Yet utilizing cold atoms here helps to minimize losses such as those induced by collisional and transit decoherence processes. In that case the experimental system can be as simple as the system in the theoretical model, these decoherence processes are negligible, and the alignment of the laser beams is more flexible.

To make a DT system as simple as possible, we put all population to a single Zeeman state. This can be achieved with strong $\sigma+$ laser fields driving the transitions from the ground states $|F=1\rangle$ and $|F=2\rangle$ to the excited states $|F'=2\rangle$ and $|F''=2\rangle$, where $F$, $F'$ and $F''$ denote the hyperfine states in the $|5S_{1/2}\rangle$, $|5P_{1/2}\rangle$ and $|5P_{3/2}\rangle$ energy levels, respectively. Hence, the Zeeman state $|F=2,m=2\rangle$ was the only dark ground state in the system. All population can be optically pumped to it. The two probe fields drove the transitions from $|F=2\rangle$ to $|F'=2\rangle$ and $|F''=2\rangle$ with the $\sigma-$ polarization. Because all the ground Zeeman states other than $|F=2,m=2\rangle$ had no population, only the probe transitions from $|F=2,m=2\rangle$ to $|F'=2,m=1\rangle$ and $|F''=2,m=1\rangle$ were relevant. Consequently, the entire atom-light coupling scheme becomes a simple DT system as shown in Fig.~1a, and all other states and transitions are irrelevant.

The probe field of $\varepsilon_A$ ($\varepsilon_B$) and the two coupling fields of $\Omega_{A1}$ and $\Omega_{A2}$ ($\Omega_{B1}$ and $\Omega_{B2}$) had wavelengths of about 795 nm (780 nm). The spontaneous decay rate of the excited states, $\Gamma$, is 2$\pi$$\times$6 MHz. Figure~1c shows the schematic experimental setup. We first overlapped $\Omega_{A1}$ and $\Omega_{B1}$, and sent them through an acousto-optic modulator (AOM) and an electro-optic modulator (EOM). The EOM generated $\Omega_{A2}$ and $\Omega_{B2}$. We set the two-photon detunings $\delta$ and $-\delta$ asymmetrically by changing the operation frequencies of EOM and AOM. 

After coming out of the EOM, the two 795 nm coupling fields and the two 780 nm ones were separated, propagated through different paths and then were overlapped again. We tuned the difference of the two path lengths to ensure that $\theta = \pi$. Propagating along the major axis of the atom cloud, the probe beams were focused to an $e^{-2}$ full width of 150 $\mu$m. The coupling beams had a much larger size to cover the entire atom cloud. We set the separation angle between the propagation directions of the probe and coupling beams to around 0.35$^{\circ}$. The input probe pulse has the Gaussian shape with an $e^{-2}$ full width of 2.5 $\mu$s and a peak power of 15 nW. Other details of the experimental setup are similar to those in Refs.~\citeR{OurPRL13} and \citeR{OurPRL12} dealing with the EIT-based slow and stationary light for $\Lambda$ or double-$\Lambda$ systems involving single atomic ground-state coherences.

\subsect{Experimental procedure and timing sequence.} Before each measurement, we first reduced the MOT repumping field intensity from 1.1 to 0.0044 mW$\cdot$cm$^{-2}$ for about 7 ms. This can increase the atomic density and, thus, the OD of the system. Then, we switched off the magnetic, repumping and trapping fields of the MOT at $t =$ -718, -74 and -48 $\mu$s, respectively ($t = 0$ denoted as the time of the input probe pulse peak). An optical pumping field was employed during the period from $t =$ -74 to -18  $\mu$s, driving the transition from $|F=1\rangle$ to $|F''=2\rangle$ with an intensity of 5 mW$\cdot$cm$^{-2}$ and the $\sigma+$ polarization. The coupling fields $\Omega_{A1}$ and $\Omega_{B1}$ with intensities of 98 and 12 mW$\cdot$cm$^{-2}$ were switched on at $t =$ -78 $\mu$s. This optical pumping field together with the coupling fields $\Omega_{A1}$ and $\Omega_{B1}$ optically pumped all population to the Zeeman state of $|F=2,m=2\rangle$. At $t =$ -5  $\mu$s, we reduced $\Omega_{A1}$ and $\Omega_{B1}$ to the designated experimental value and switched on $\Omega_{A2}$ and $\Omega_{B2}$ with the same value. After the four coupling fields for the SSL were present, we fired the probe pulse $\varepsilon_A$. Two photo multipliers (Hamamatsu PMT H6780-20 and H10720) were employed to detect the signals of the two output probe pulses $\varepsilon_A$ and $\varepsilon_B$. Signals from the PMTs were sent to a digital oscilloscope (Agilent MSO6014A). Data were averaged 200 times by the oscilloscope before being acquired by a computer. After the measurement was complete, we turned off the coupling fields and turned back on the MOT. The above measurement sequence was repeated every 0.15 s.

\subsect{Determination of experimental parameters.}
With two single-$\Lambda$ and two double-$\Lambda$ measurements, we determined the experimental parameters of the optical density ($\alpha$), coupling Rabi frequencies ($\Omega_{A1}$, $\Omega_{A2}$, $\Omega_{B1}$ and $\Omega_{B2}$), dephasing rates of the ground-state coherences ($\gamma_1$ and $\gamma_2$) and degree of phase mismatch ($\Delta_k L$). The procedure is described below. We first measured the slow light output in the single-$\Lambda$ system formed by $|0\rangle$, $|1\rangle$ and $|A\rangle$. During the measurement, the coupling field of $\Omega_{A1}$ was constantly present. The experimental data are shown in Fig.~5a. Varying the parameters of $\alpha$, $\Omega_{A1}$ and $\gamma_1$ in the numerical calculation, we fitted the data with the theoretical prediction. Note that $\Delta_k$ plays no role in the output amplitude of the probe pulse in the single-$\Lambda$ system. The best fit determined the experimental parameters of $\alpha$, $\Omega_{A1}$ and $\gamma_1$. 

\FigFive

Similarly, we measured the slow light output in another EIT system formed by $|0\rangle$, $|2\rangle$ and $|A\rangle$ to determine $\alpha$, $\Omega_{A2}$ and $\gamma_2$. The experimental data are shown in Fig.~5c. We imposed a constraint that $\alpha$ values in the two EIT systems can only differ within the uncertainty. The determined $\gamma_2$ is rather large as compared with the dephasing rate in our previous work\citeS{OurPRL13}. This $\gamma_2$ of 3.7$\times$10$^{-3}$$\Gamma$ corresponds to a coherence time of 3.6 $\mu$s. Because the data of light storage in this work still showed a reasonably good coherence time of 76 $\mu$s as compared with the previous work, this large $\gamma_2$ might be due to the new fiber-based electro-optic modulator (EOM) used in the experiment. The EOM produced the sidebands which served as the coupling fields of $\Omega_{A2}$ and $\Omega_{B2}$ (see Fig.~1c). 

After $\alpha$, $\Omega_{A1}$ and $\gamma_1$ were determined, we kept $\Omega_{A1}$ and the experimental condition unchanged, and performed the measurement of the double-$\Lambda$ system formed by $|0\rangle$, $|1\rangle$, $|A\rangle$ and $|B\rangle$. With $\varepsilon_A$ being the only input, Fig.~5b shows the experimental data of the two output probe pulses. We fitted the data with the predictions, and the best fit determined $\Omega_{B1}$ and $\Delta_k L$. Note that the same product of $\Delta_k$ and $L$ results in the same calculation result. Similarly, we measured two output probe pulses in another double-$\Lambda$ system formed by $|0\rangle$, $|2\rangle$, $|A\rangle$ and $|B\rangle$ to determine $\Omega_{B2}$ and $\Delta_k L$. The experimental data are shown in Fig.~5d. The $(\Delta_k L)$'s determined from the two double-$\Lambda$ systems were about the same. The experimental condition was set and determined by the above procedure. We kept the condition unchanged and measured the data of the DT system. The determined parameters were used to calculate the theoretical predictions as shown in Figs.~2a, 2b and 2c.



\sect{Acknowledgments} \\ \noindent
This work was supported by the Ministry of Science and Technology of Taiwan under Grant Nos. 101-2112-M-007-008-MY3, 103-2923-M-007-001 and 103-2119-M-007-011, by the project TAP LLT 01/2012 of the Research Council of Lithuania and the Ministry of Science and Technology of Taiwan, and by the EU FP7 IRSES project COLIMA (contract PIRSES-GA-2009-247475).

\sect{Author contributions} \\ \noindent
I.A.Y., G.J., J.R., V.K., and H-W.C. conceived the experiment. I.A.Y. and M-J.L. designed the experimental setup and methods. M-J.L., C-Y.L., K-F.C. and H-W.C. built the setup and carried out the experiment. M-J.L. and C-Y.L. calculated the predictions and analyzed the data under the supervision of I.A.Y. The manuscript was written by I.A.Y. with the help from G.J., M-J.L., J.R., C-Y.L. and V.K.

\sect{Additional information} \\ \noindent
The authors declare no competing financial interests. 

\vspace*{2\baselineskip}
{\center{\Large{\bf 
Supplementary Notes \\
}}}
\vspace*{-0.5\baselineskip}

\renewcommand{\theequation}{S\arabic{equation}}
\setcounter{equation}{0}
\renewcommand{\thefigure}{S\arabic{figure}}
\renewcommand{\thesection}{{\normalsize\arabic{section}.\hspace*{-0.45cm}}}
\newcommand{\SNsection}[1]{{\raggedright\section{{\normalsize #1}}}}
\newcommand{\figDL}[1]{S1#1}
\newcommand{\FigDL}[1]{	
	\begin{figure}[#1]
	{\center\includegraphics[width=5.5cm]{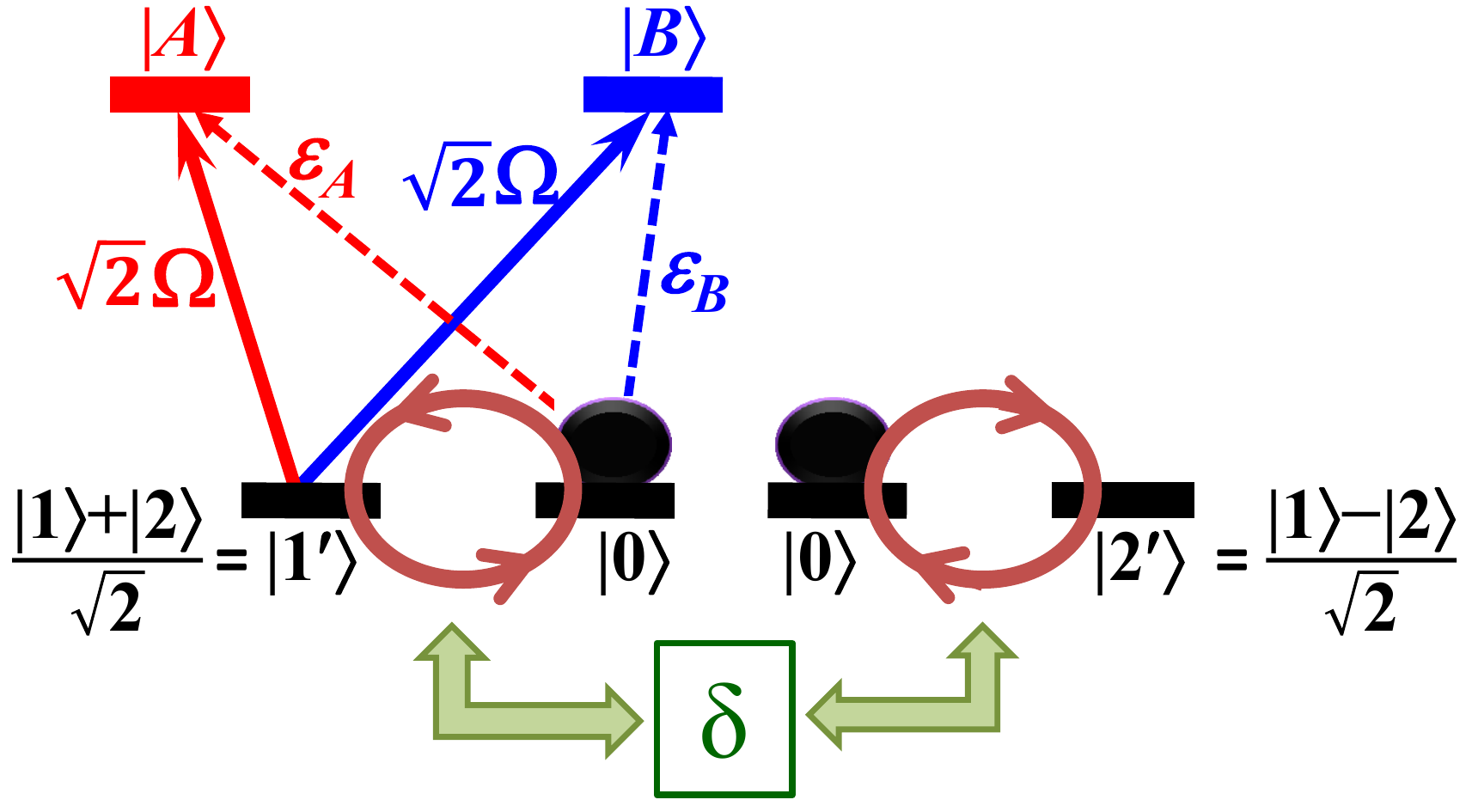}\\}
	\textbf{Figure \figDL{} $\vert$ 
Transition diagram equivalent to the DT system at $\theta = 0$.
	}
It consists of a double-$\Lambda$ system and a two-ground-state system which couple with each other via the ground-state coherences.
	\end{figure}
}

\newcommand{\figOsc}[1]{S2#1}
\newcommand{\FigOsc}[1]{	
	\begin{figure}[#1]
	{\center\includegraphics[width=8.8cm]{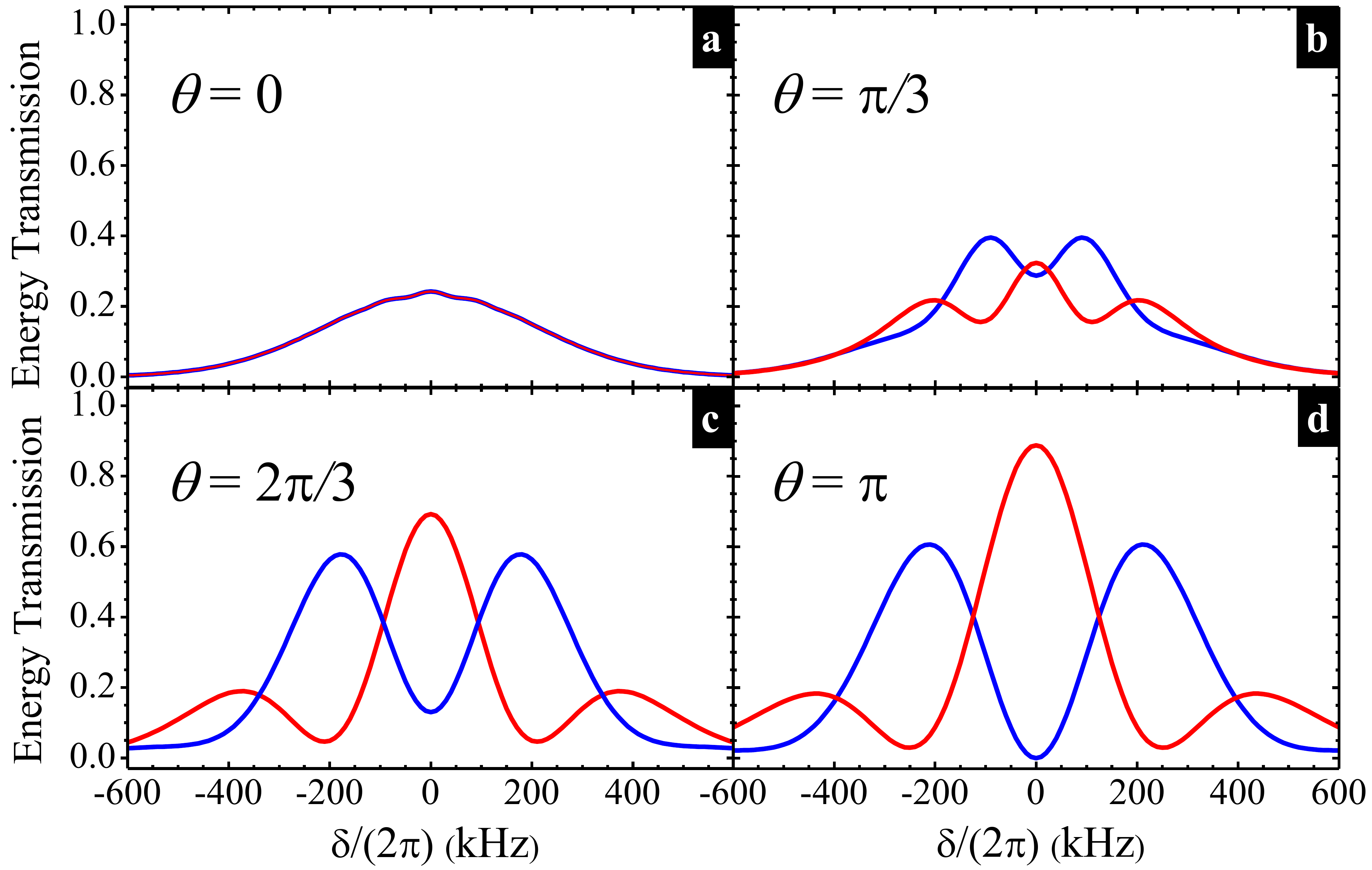}\\}
	\textbf{Figure \figOsc{} $\vert$ 
Theoretical predictions of the probe energy transmissions versus the detuning at different values of $\theta$.
	}
Red and blue lines are the transmissions of $\varepsilon_A$ and  $\varepsilon_B$, respectively. In all the plots we set the calculation parameters to the experimental condition that $\alpha$ = 20, $|\Omega_{A1}|$ = $|\Omega_{A2}|$ = $|\Omega_{B1}|$ = $|\Omega_{B2}|$ = 0.51$\Gamma$, and only the probe pulse $\varepsilon_A$ with the $e^{-2}$ full width of 2.5 $\mu$s or 94$\Gamma^{-1}$ is present in the input. The phase mismatch and ground-state coherence dephasing are not included in the calculation. In \textbf{a-d}, $\theta$ = 0, $\pi/3$, $2\pi/3$ and $\pi$, respectively. The two lines completely overlap in \textbf{a}.
	\end{figure}
}
\newcommand{\figTheta}[1]{S3#1}
\newcommand{\FigTheta}[1]{	
	\begin{figure}[#1]
	{\center\includegraphics[width=6cm]{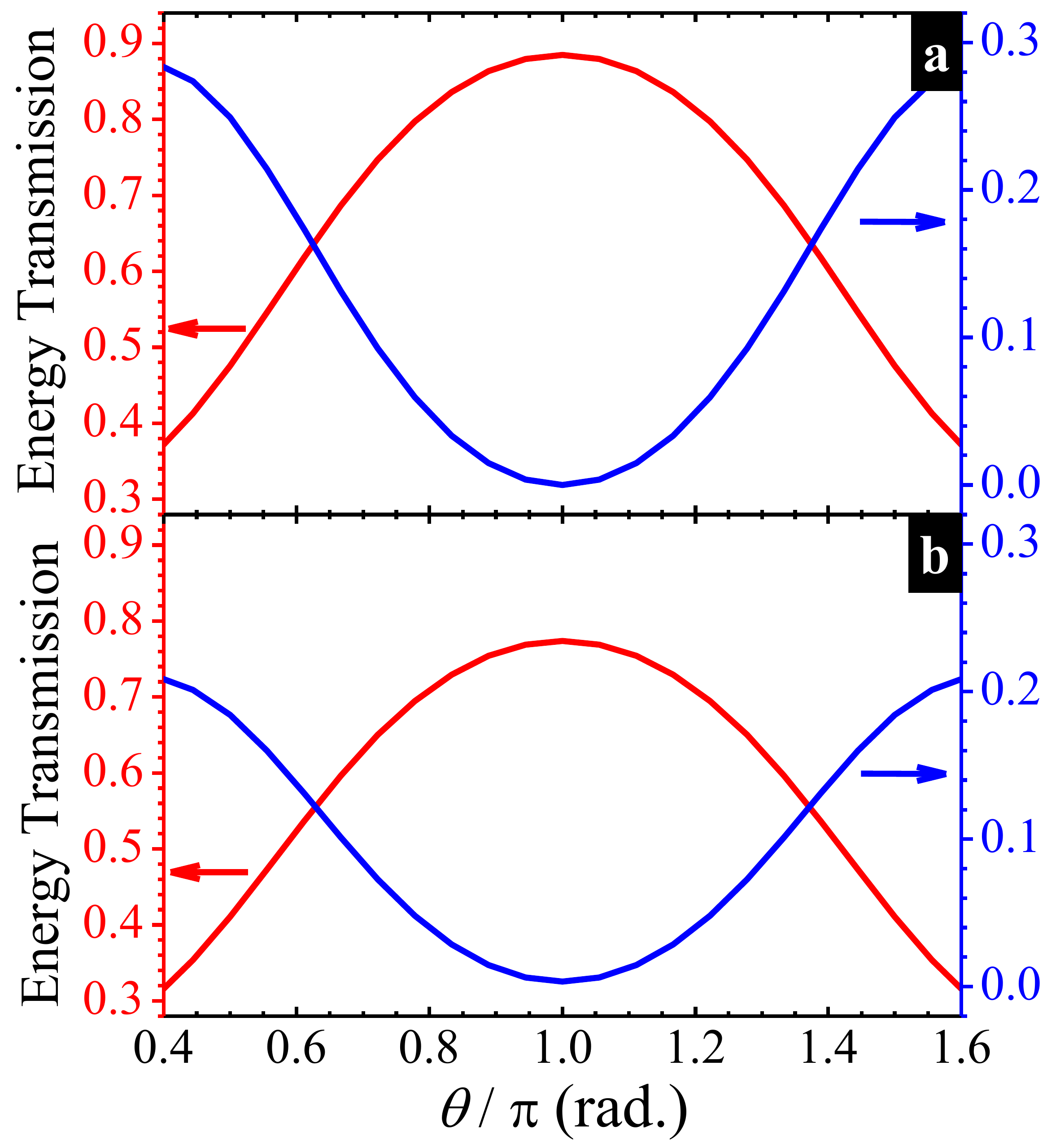}\\}
	\textbf{Figure \figTheta{} $\vert$ 
Theoretical predictions of the probe energy transmissions versus $\theta$ at the zero detuning.
	}
Red and blue lines are the transmissions of  $\varepsilon_A$ and  $\varepsilon_B$, respectively. The calculation parameters of OD, coupling Rabi frequencies and input probe pulse are the same as those in Fig.~\figOsc{}.
	\textbf{a,}
The phase mismatch and ground-state coherence dephasing are not included.
	\textbf{b,}
We take $\Delta_k L$ = 0.6, $\gamma_1$ = 0 and $\gamma_2$ = $3.7\times10^{-3}$$\Gamma$ corresponding to the experimental situation.
	\end{figure}
}
\newcommand{\figDelta}[1]{S4#1}
\newcommand{\FigDelta}[1]{	
	\begin{figure}[#1]
	{\center\includegraphics[width=8.8cm]{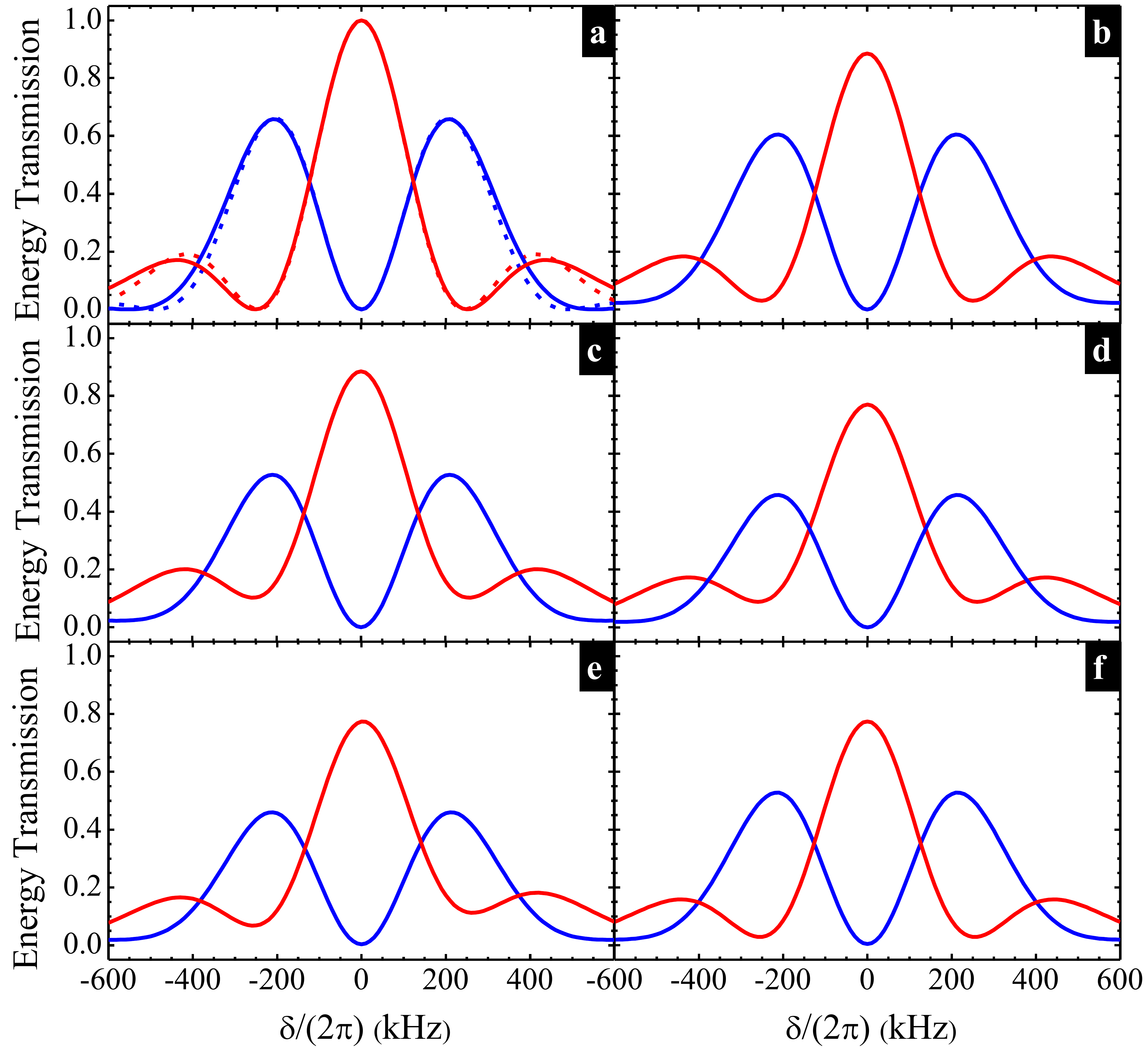}\\}
	\textbf{Figure \figDelta{} $\vert$ 
Theoretical predictions of the probe energy transmissions versus the detuning at $\theta = \pi$.
	}
Red and blue lines are the transmissions of $\varepsilon_A$ and  $\varepsilon_B$, respectively. Solid lines are the numerical predictions and dashed lines are the analytical results given by Eq.~(6) of the main text. The calculation parameters of OD and coupling Rabi frequencies are the same as those in Fig.~\figOsc{}. Only probe $\varepsilon_A$ is present in the input. In \textbf{a}, $\varepsilon_A$ is a continuous wave; in \textbf{b-f}, $\varepsilon_A$ is a pulse the same as that in Fig.~\figOsc{}.
	\textbf{a,b,}
$\Delta_k L$ = 0 and $\gamma_1$ = $\gamma_2$ = 0.
	\textbf{c,}
$\Delta_k L$ = 0.6 and $\gamma_1$ = $\gamma_2$ = 0.
	\textbf{d,}
$\Delta_k L$ = 0.6 and $\gamma_1$ = $\gamma_2$ = $1.85\times10^{-3}$$\Gamma$.
	\textbf{e,}
$\Delta_k L$ = 0.6, $\gamma_1$ = 0 and $\gamma_2$ = $3.7\times10^{-3}$$\Gamma$.
	\textbf{f,}
$\Delta_k L$ = 0, $\gamma_1$ = 0 and $\gamma_2$ = $3.7\times10^{-3}$$\Gamma$.
	\end{figure}
}

\SNsection{Analytical Solution of Continuous-Wave Spinor Slow Light}
\vspace*{-0.5\baselineskip}

We consider the spinor slow light (SSL) forming when two probe beams are coupled with two atomic coherences in the double-tripod (DT) scheme of atom-light coupling. The time evolution of the SSL is described by the following equations: 

\begin{eqnarray}
	\frac{1}{c}\frac{\partial}{\partial t} \VT{\varepsilon_A}{\varepsilon_B}
		+\frac{\partial}{\partial z} \VT{\varepsilon_A}{\varepsilon_B}
		+ \MT{i\Delta_k/2}{0}{0}{-i\Delta_k/2} \VT{\varepsilon_A}{\varepsilon_B}
		\nonumber \\
		=i\frac{\alpha\Gamma}{2L} \VT{\rho_A}{\rho_B},
\label{eq:MSE}
		\\
	\frac{\partial}{\partial t} \VT{\rho_A}{\rho_B} 
		= \frac{i}{2} \VT{\varepsilon_A}{\varepsilon_B} 
		+\frac{i}{2} \MT{\Omega_{A1}}{\Omega_{A2}}{\Omega_{B1}}{\Omega_{B2}}
		\VT{\rho_1}{\rho_2}
		\nonumber \\
		-\frac{\Gamma}{2} \VT{\rho_A}{\rho_B},
\label{eq:OBE1}
		\\
	\frac{\partial}{\partial t} \VT{\rho_1}{\rho_2} 
		=\frac{i}{2} 
		\MT{\Omega_{A1}^*}{\Omega_{B1}^*}{\Omega_{A2}^*}{\Omega_{B2}^*} 
		\VT{\rho_A}{\rho_B} 
		+\MT{i\delta}{0}{0}{-i\delta} \VT{\rho_1}{\rho_2}
		\nonumber \\
		+\MT{-\gamma_1}{0}{0}{-\gamma_2} \VT{\rho_1}{\rho_2},
\label{eq:OBE2}
\end{eqnarray}
where $\varepsilon_A$, $\varepsilon_B$, $\Omega_{A1}$, $\Omega_{A2}$, $\Omega_{B1}$ and $\Omega_{B2}$ are the Rabi frequencies of the probe and coupling fields driving the atomic transitions depicted in Fig.~1a of the main text, $\rho_{A}$ (or $\rho_{B}$) is the coherence of the probe transition $|0\rangle \rightarrow |A\rangle$ (or $|0\rangle \rightarrow |B\rangle$), $\rho_1$ (or $\rho_2$) is the atomic ground-state coherence between $|0\rangle$ and $|1\rangle$ (or $|2\rangle$), $\Gamma$ is the spontaneous decay rate equal to $2\pi$$\times$6 MHz in our experiment, $\alpha$ and $L$ are the optical density (OD) and length of the medium, $\delta$ is the two-photon detuning as illustrated in Fig.~1a of the main text, $\gamma_1$ (or $\gamma_2$) is the dephasing rate of the atomic coherence $\rho_1$ (or $\rho_2$), and $\Delta_k \equiv (-{\bf\vec{k}}_{pA} +{\bf\vec{k}}_{cA} -{\bf\vec{k}}_{cB} +{\bf\vec{k}}_{pB}) \cdot {\bf\hat{z}}$ describes the effect of the phase mismatch. In the definition of $\Delta_k$,  ${\bf\vec{k}}_{pA}$ and ${\bf\vec{k}}_{pB}$ are the wave vectors of the probe fields $\varepsilon_A$ and $\varepsilon_B$, and ${\bf\vec{k}}_{cA}$ and ${\bf\vec{k}}_{cB}$ are those of the coupling fields $\Omega_{A1}$ and $\Omega_{B1}$ (or $\Omega_{A2}$ and $\Omega_{B2}$). The analytical expressions and numerical predictions in the paper and in the supplement are based on the above equations. 

We focus on the case where the complex Rabi frequencies of the four coupling fields have the same amplitude of $\Omega$. Thus, the complex Rabi frequency of the $n$th coupling field is $\Omega_n = \Omega e^{i\theta_n}$, with $n = A1$, $A2$, $B1$ or $B2$, where $\theta_n$ is the phase of coupling field. We define $\theta \equiv (\theta_{A1}-\theta_{A2})-(\theta _{B1}-\theta _{B2})$ to be a relative phase among the four coupling fields. To simplify the analytical derivation, we take $\gamma_1$ = 0 = $\gamma_2$ and $\Delta_k$ = 0. For the continuous waves, Eqs.~(\ref{eq:MSE})-(\ref{eq:OBE2}) reduce to
\begin{eqnarray}
	\frac{\partial}{\partial z} \VT{\varepsilon_A}{\varepsilon_B}
		= i\frac{\alpha\Gamma}{2L} \VT{\rho_A}{\rho_B},
\label{eq:MSEss}
	\\
	0 = \frac{i}{2} \VT{\varepsilon_A}{\varepsilon_B} 
		+\frac{i}{2} \MT{ i\Omega}{\Omega}{\Omega}{i \Omega}
		\VT{\rho_1}{\rho_2}
		-\frac{\Gamma}{2} \VT{\rho_A}{\rho_B},
\label{eq:OBE1ss}
	\\
	0 = \frac{i}{2} \MT{-i\Omega}{\Omega}{\Omega}{-i\Omega} 
		\VT{\rho_A}{\rho_B} 
		+\MT{i\delta}{0}{0}{-i\delta} \VT{\rho_1}{\rho_2}.
\label{eq:OBE2ss}
\end{eqnarray}
Here we set the phases of individual coupling fields ($\theta_{A1}$, $\theta_{A2}$, $\theta_{B1}$, $\theta_{B2}$) = ($\pi/2$, 0, 0, $\pi/2$) such that their relative phase is $\theta = \pi$. Other combinations, such as ($\pi/2$, $-\pi/2$, 0, 0), ($\pi$, 0, 0, 0), (0, 0, 0, $\pi$), etc., can also achieve the same result shown at the end of this section. To derive the solution of the above equations, we first make use of Eqs.~(\ref{eq:OBE1ss}) and (\ref{eq:OBE2ss}) to eliminate $(\rho_1,\rho_2)$, subsequently expressing $(\rho_A,\rho_B)$ in terms of $(\varepsilon_A,\varepsilon_B)$:
\begin{equation}
	 \VT{\rho_A}{\rho_B} 
		=\frac{i}{\Gamma}\MT{1}{-\beta}{\beta}{1}^{-1}
		\VT{\varepsilon_A}{\varepsilon_B},
\end{equation}
where
\begin{equation}
		\beta \equiv \frac{\Omega^2}{\delta\Gamma}.
\end{equation}
Consequently Eq.~(\ref{eq:MSEss}) becomes
\begin{equation}
	\frac{\partial}{\partial z} \VT{\varepsilon_A}{\varepsilon_B}
		=-\frac{\alpha}{2L} \frac{1}{1+\beta^2}
		\MT{1}{\beta}{-\beta}{1} \VT{\varepsilon_A}{\varepsilon_B}.
\label{eq:CoupledEoM}
\end{equation}
The matrix on the right-hand side of the above equation can be diagonalized by transforming the two probe fields $(\varepsilon_A,\varepsilon_B)$ to new variables $(\varepsilon_a,\varepsilon_b)$ via a unitary transformation
\begin{equation}
	 \VT{\varepsilon_a}{\varepsilon_b} = \frac{1}{\sqrt{2}}
		\MT{1}{i}{i}{1} \VT{\varepsilon_A}{\varepsilon_B}.
\label{eq:Transform}
\end{equation} 
Based on Eq.~(\ref{eq:CoupledEoM}), the propagation equation for $(\varepsilon_a,\varepsilon_b)$ and its solution are
\begin{equation}
	\frac{\partial}{\partial z} \VT{\varepsilon_a}{\varepsilon_b}
		=-\frac{\alpha}{2L} \frac{1}{1+\beta^2}
		\MT{1-i\beta}{0}{0}{1+i\beta} \VT{\varepsilon_a}{\varepsilon_b}
\end{equation}
and
\begin{eqnarray}
	\VT{\varepsilon_a(L)}{\varepsilon_b(L)} 
		&& = \exp\left(-\frac{\alpha}{2}\frac{1}{1+\beta^2}\right)
		\nonumber \\
	\times &&
		\MT{\exp\left(i\frac{\alpha}{2}\frac{\beta}{1+\beta^2}\right)}{0}
		{0}{\exp\left(-i\frac{\alpha}{2}\frac{\beta}{1+\beta^2}\right)} 
		\VT{\varepsilon_a(0)}{\varepsilon_b(0)},
	\hspace*{1cm}
\end{eqnarray}
respectively. The transformed fields $\varepsilon_a$ and $\varepsilon_b$ represent the two normal modes inside the medium. Using the transformation Eq.~(\ref{eq:Transform}), we obtain the output probe fields $\varepsilon_A$ and $\varepsilon_B$ as a function of their inputs:
\begin{eqnarray}
	\VT{\varepsilon_A(L)}{\varepsilon_B(L)} && 
		= \exp\left(-\frac{\alpha}{2} \frac{1}{1+\beta^2}\right) 
		\nonumber \\
	&& \times
		\MT{\cos\left(\frac{\alpha}{2}\frac{\beta}{1+\beta^2}\right)}
		{-\sin\left(\frac{\alpha}{2}\frac{\beta}{1+\beta^2}\right)}
		{\sin\left(\frac{\alpha}{2}\frac{\beta}{1+\beta^2}\right)}
		{\cos\left(\frac{\alpha}{2}\frac{\beta}{1+\beta^2}\right)} 
		\VT{\varepsilon_A(0)}{\varepsilon_B(0)}.
	\hspace*{1cm}
\label{eq:OscillationExact}
\end{eqnarray}
For $\beta \gg 1$ or $\Omega^2 \gg \delta\Gamma$, the solution simplifies to
\begin{equation}
	\VT{\varepsilon_A(L)}{\varepsilon_B(L)} = 
		\exp\left(-\frac{2\phi^2}{\alpha}\right)
		\MT{\cos\phi}{-\sin\phi}{\sin\phi}{\cos\phi}
		\VT{\varepsilon_A(0)}{\varepsilon_B(0)},
\label{eq:Oscillation}
\end{equation}
where
\begin{equation}
		\phi = \frac{\alpha}{2}\frac{\delta\Gamma}{\Omega^2}. 
\label{eq:Phase}
\end{equation}
The above two equations represent Eqs.~(4) and (5) of the main text.

\vspace*{\baselineskip}
\SNsection{Double-Tripod System Equivalent to Two Coupled $\Lambda$ Systems}
\vspace*{-0.5\baselineskip}

Considering that the four coupling fields having the same amplitude, their relative phase $\theta$ equal to $\pi$, the dephasing rates and phase mismatch being negligible, Eqs.~(\ref{eq:OBE1}) and (\ref{eq:OBE2}) become
\begin{equation}
	\frac{\partial}{\partial t} \VT{\rho_A}{\rho_B} 
		= \frac{i}{2} \VT{\varepsilon_A}{\varepsilon_B} 
		+\frac{i}{2} \MT{\Omega}{\Omega}{\Omega}{-\Omega}
		\VT{\rho_1}{\rho_2}
		-\frac{\Gamma}{2} \VT{\rho_A}{\rho_B},
\label{eq:OBE1atPi}
\end{equation}
\begin{equation}
	\frac{\partial}{\partial t} \VT{\rho_1}{\rho_2} 
		=\frac{i}{2} \MT{\Omega}{\Omega}{\Omega}{-\Omega}
		\VT{\rho_A}{\rho_B} 
		+\MT{i\delta}{0}{0}{-i\delta} \VT{\rho_1}{\rho_2}.
\label{eq:OBE2atPi}
\end{equation}
For convenience here we set ($\theta_{A1}$, $\theta_{A2}$, $\theta_{B1}$, $\theta_{B2}$) = (0, 0, 0, $\pi$). Other combinations, such as ($\pi/2$, 0, 0, $\pi/2$), ($\pi/2$, $-\pi/2$, 0, 0), ($\pi$, 0, 0, 0), etc., can also achieve the same conclusion shown at the end of this section. We introduce a unitary transformation $U$ given by
\begin{equation}
	U = \frac{1}{\sqrt{2}}\MT{1}{1}{1}{-1}.
\end{equation}
and insert it in Eqs.~(\ref{eq:OBE1atPi}) and (\ref{eq:OBE2atPi}) in the following way:
\begin{eqnarray}
	\frac{\partial}{\partial t} \VT{\rho_A}{\rho_B} 
		&& = \frac{i}{2} \VT{\varepsilon_A}{\varepsilon_B} 
		\nonumber \\
	&& +\frac{i}{2} \left( \MT{ \Omega}{ \Omega}{ \Omega}{- \Omega} U^{-1} \right)
		\left( U \VT{\rho_1}{\rho_2} \right) -\frac{\Gamma}{2} \VT{\rho_A}{\rho_B},
	\hspace*{1cm}
\label{eq:OBE1withU}
\end{eqnarray}
\vspace*{-\baselineskip}
\begin{eqnarray}
	\frac{\partial}{\partial t} \left( U \VT{\rho_1}{\rho_2} \right)
		&& =\frac{i}{2} 
		\left( U \MT{ \Omega}{ \Omega}{ \Omega}{- \Omega} \right) 
		\VT{\rho_A}{\rho_B} 
		\nonumber \\
	&& + \left( U \MT{i\delta}{0}{0}{-i\delta} U^{-1} \right) \left( U \VT{\rho_1}{\rho_2} \right). 
	\hspace*{1.6cm}
\label{eq:OBE2withU}
\end{eqnarray}
It is convenient to define two new variables $\rho_{1'}$ and $\rho_{2'}$ as
\begin{equation}
	\VT{\rho_{1'}}{\rho_{2'}} = U \VT{\rho_1}{\rho_2}.
\label{eq:TransformGroundRho}
\end{equation}
The new variable $\rho_{1'}$ (or $\rho_{2'}$) represents the ground-state coherence between the states $|1'\rangle$ (or $|2'\rangle$) and $|0\rangle$, where $|1'\rangle$ and $|2'\rangle$ are the following superpositions of the original ground states $|1\rangle$ and $|2\rangle$:
\begin{eqnarray}
	|1'\rangle = \frac{1}{\sqrt{2}} (|1\rangle + |2\rangle), \\
	|2'\rangle = \frac{1}{\sqrt{2}} (|1\rangle - |2\rangle).
\end{eqnarray}
Finally, Eqs.~(\ref{eq:OBE1withU}) and (\ref{eq:OBE2withU}) become 
\begin{eqnarray}
	\frac{\partial}{\partial t} \VT{\rho_A}{\rho_B} 
		= \frac{i}{2} \VT{\varepsilon_A}{\varepsilon_B} 
		+\frac{i}{2} \MT{\sqrt{2}\Omega}{0}{0}{\sqrt{2}\Omega}
		\VT{\rho_{1'}}{\rho_{2'}}
		\nonumber \\
	-\frac{\Gamma}{2} \VT{\rho_A}{\rho_B},
\label{eq:OBE1equiv}
\end{eqnarray}
\begin{equation}
	\frac{\partial}{\partial t} \VT{\rho_{1'}}{\rho_{2'}}
		=\frac{i}{2} \MT{\sqrt{2}\Omega}{0}{0}{\sqrt{2}\Omega}
		\VT{\rho_A}{\rho_B} 
		+\MT{0}{i\delta}{i\delta}{0} \VT{\rho_{1'}}{\rho_{2'}}.
\label{eq:OBE2equiv}
\end{equation}
According to the above equations, there are two EIT systems. One consists of $\varepsilon_A$, $\rho_A$, and $\rho_{1'}$, the other being made of $\varepsilon_B$, $\rho_B$, and $\rho_{2'}$. The coupling fields in both systems have the same Rabi frequency of $\sqrt{2}\Omega$.  In Eqs.~(\ref{eq:OBE1equiv}) and (\ref{eq:OBE2equiv}), the only matrix with non-zero off-diagonal terms is
\[
	\MT{0}{i\delta}{i\delta}{0}.
\]
It indicates that the interaction between the two ground-state coherences $\rho_{1'}$ and $\rho_{2'}$ (i.e. the coupling between the two EIT systems) is induced by the detuning $\delta$. Therefore, the DT system is equivalent to the two coupled EIT systems as depicted in Fig.~1b of the main text.

\vspace*{\baselineskip}
\SNsection{Degenerate Double-Tripod System}
\vspace*{-0.5\baselineskip}

The DT system becomes degenerate as the pair of the coupling fields in each tripod have the same complex Rabi frequency, i.e. the relative phase $\theta = 0$. Under this special condition, the DT system will be shown to be equivalent to a double-$\Lambda$ system and no oscillation can occur between the two probe fields. For convenience but without loss of generality, the four coupling fields have the same complex Rabi frequency of $\Omega$. Considering the dephasing rates and phase mismatch to be negligible, Eqs.~(\ref{eq:OBE1}) and (\ref{eq:OBE2}) become
\begin{eqnarray}
	\frac{\partial}{\partial t} \VT{\rho_A}{\rho_B} 
		= \frac{i}{2} \VT{\varepsilon_A}{\varepsilon_B} 
		+\frac{i}{2} \MT{\Omega}{\Omega}{\Omega}{\Omega}
		\VT{\rho_1}{\rho_2}
		-\frac{\Gamma}{2} \VT{\rho_A}{\rho_B},
		\\
	\frac{\partial}{\partial t} \VT{\rho_1}{\rho_2} 
		=\frac{i}{2} \MT{\Omega}{\Omega}{\Omega}{\Omega}
		\VT{\rho_A}{\rho_B} 
		+\MT{i\delta}{0}{0}{-i\delta} \VT{\rho_1}{\rho_2}.
\end{eqnarray}
Using the two variables $\rho_{1'}$ and $\rho_{2'}$ defined in Eq.~(\ref{eq:TransformGroundRho}), we rewrite the above equations as 
\begin{eqnarray}
	\frac{\partial}{\partial t} \VT{\rho_A}{\rho_B} 
		= \frac{i}{2} \VT{\varepsilon_A}{\varepsilon_B} 
		+\frac{i}{2} \MT{\sqrt{2}\Omega}{0}{\sqrt{2}\Omega}{0}
		\VT{\rho_{1'}}{\rho_{2'}}
		\nonumber \\
	-\frac{\Gamma}{2} \VT{\rho_A}{\rho_B},
\label{eq:OBE1DL}
\end{eqnarray}
\begin{equation}
	\frac{\partial}{\partial t} \VT{\rho_{1'}}{\rho_{2'}} 
		=\frac{i}{2} \MT{\sqrt{2}\Omega}{\sqrt{2}\Omega}{0}{0}
		\VT{\rho_A}{\rho_B} 
		+\MT{0}{i\delta}{i\delta}{0} \VT{\rho_{1'}}{\rho_{2'}}.
\label{eq:OBE2DL}
\end{equation}
In Eq.~(\ref{eq:OBE1DL}), both $\rho_A$ and $\rho_B$ (or $\varepsilon_A$ and $\varepsilon_B$) couple to the same ground-state coherence $\rho_{1'}$ as indicated by the term
\begin{displaymath}
	\frac{i}{2} \MT{\sqrt{2}\Omega}{0}{\sqrt{2}\Omega}{0} \VT{\rho_{1'}}{\rho_{2'}}.
\end{displaymath}
In  Eq.~(\ref{eq:OBE2DL}), the ground-state coherence $\rho_{2'}$ does not interact with $\rho_A$ and $\rho_B$ directly as indicated by the term
\begin{displaymath}
	\frac{i}{2} \MT{\sqrt{2}\Omega}{\sqrt{2}\Omega}{0}{0} \VT{\rho_A}{\rho_B},
\end{displaymath}
but the two ground-states coherences are coupled via the detuning $\delta$ as indicated by the term
\begin{displaymath}
	\MT{0}{i\delta}{i\delta}{0} \VT{\rho_{1'}}{\rho_{2'}}.
\end{displaymath}
Therefore, the DT system at $\theta = 0$ can be decomposed into two coupled subsystems. One is the double-$\Lambda$ system in which the two probe fields share the common coherence $\rho_{1'}$. The other is the system only consisting of two ground states in which no light appears but the coherence $\rho_{2'}$ exists. Figure~\figDL{} depicts the transition diagram equivalent to the DT system at $\theta = 0$. Because the two probe fields share the common ground-state coherence and interact with the medium in a similar way, their outputs always behave the same which will be demonstrated in the next section. 

\FigDL{t}

\vspace*{\baselineskip}
\SNsection{Oscillation Behaviors at Different Relative Phases of the Coupling Fields}
\vspace*{-0.5\baselineskip}

\FigOsc{t}

The SSL oscillations can also be observed for the relative phase $\theta$ of the coupling fields other than $\pi$. Increasing the deviation of $\theta$ from $\pi$ makes the oscillation less prominent, as the temporal profile of the probe field is a pulse. Finally, for a maximum deviation at $\theta = 0$, the DT system becomes equivalent to the double-$\Lambda$ system as discussed in the previous section, and no oscillation can occur between the probe fields. 

Figures~\figOsc{a}-\figOsc{d} show the theoretical predictions of the probe energy transmission as a function of the detuning $\delta$ at different values of $\theta$. The predictions were numerically calculated from Eqs.~(\ref{eq:MSE})-(\ref{eq:OBE2}) by taking the parameters of input probe pulse width, optical density and coupling Rabi frequencies used in the experiment. The phase mismatch $\Delta_k$ and the ground-state dephasing rates $\gamma_1$ and $\gamma_2$ are set to zero in the calculation. The prediction in Fig.~\figOsc{a} shows the two probe pulses always have the same output energy (also temporal shape) at $\theta =0$. This is the expected outcome of the double-$\Lambda$ system with the same coupling Rabi frequency in each of the two constituent $\Lambda$ systems. As demonstrated by Fig.~\figOsc{b}-\figOsc{d}, the condition of $\theta$ = $\pi$ gives the maximum contrast or difference between two output probe fields around $\delta = 0$. For this reason, $\theta$ = $\pi$ was chosen in the experiment.

\vspace*{\baselineskip}
\SNsection{Method of Setting the Relative Phase of the Coupling Fields to $\pi$}
\vspace*{-0.5\baselineskip}

\FigTheta{t}

The relative phase of the coupling fields $\theta$ can be made equal to $\pi$ directly, when the probe field is a pulse instead of a continuous wave. The theoretical predictions of the two output probe energy transmissions versus $\theta$ at the detuning $\delta = 0$ shown in Figs.~\figTheta{a} and \figTheta{b} illustrate the idea. They were numerically calculated from Eqs.~(\ref{eq:MSE})-(\ref{eq:OBE2}) with the parameters of input probe pulse width, optical density and coupling Rabi frequencies used in the experiment. As only the probe $\varepsilon_A$ is present at the input of the medium, the probe $\varepsilon_B$ will not be generated at the output if $\theta = \pi$. A larger deviation from $\pi$ in $\theta$ causes larger output energy of $\varepsilon_B$. The inclusion of the phase mismatch and ground-state coherence dephasing existing in the experimental system do not change the behavior of $\varepsilon_B$'s output energy versus $\theta$.

Experimentally, we moved the position of the prism shown in Fig.~1c of the main text and measured $\varepsilon_B$'s output energy at $\delta = 0$. According to its definition, $\theta$ is the phase difference between the beat note of $\Omega_{A1}$ plus $\Omega_{A2}$ and that of $\Omega_{B1}$ plus $\Omega_{B2}$. The frequency difference between $\Omega_{A1}$ and $\Omega_{A2}$ (or between $\Omega_{B1}$ and $\Omega_{B2}$) is about 6.8 GHz, corresponding to the beat-note wavelength of 4.4 cm. The prism position can change the optical path length of $\Omega_{A1}$ plus $\Omega_{A2}$ and, thus, adjust $\theta$. By minimizing $\varepsilon_B$'s output, we were able to properly set $\theta = \pi$.

\FigDelta{t}

\vspace*{\baselineskip}
\SNsection{Nonzero Minima and Asymmetry in the Oscillation Phenomenon}
\vspace*{-0.5\baselineskip}

Let us now discuss the discrepancy between the observed SSL output and the oscillation phenomenon predicted by Eq.~(6) of the main text. There are two major features in the discrepancy. The minima of the probe field $\varepsilon_A$ are not completely zero, and the probe transmissions are asymmetric at the positive and negative detunings. Figure~\figDelta{a} shows the predictions numerically calculated from Eqs.~(\ref{eq:MSE})-(\ref{eq:OBE2}) for a continuous-wave input probe. The numerical predictions are in a good agreement with the analytical expression given by Eq.~(6) of the main text except at large $|\delta|$ where the condition $\Omega^2 \gg \delta\Gamma$ is not held. When the input probe is a Gaussian pulse with the width used in the experiment, Fig.~\figDelta{b} clearly shows $\varepsilon_A$'s minima become about 3\%. Figures~\figDelta{c}-\figDelta{f} all consider the input probe is a pulse. When the amount of $\Delta_k L$ existing in the experiment is added to the calculation, the total output energy is affected very little but $\varepsilon_A$'s minima further increase to 10\% as shown in Fig.~\figDelta{c}. Hence, the nonzero minima are caused by the existence of the phase mismatch $\Delta_k$ and the finite frequency bandwidth of the input probe pulse. 

We further include the ground-state coherence dephasing in the calculation. As the two dephasing rates $\gamma_1$ and $\gamma_2$ of the coherences $\rho_1$ and $\rho_2$ are the same, Fig.~\figDelta{d} shows the total output energy is significantly decreased. So far, the spectra are all symmetric with respect to $\delta = 0$. By setting $\gamma_1 \neq \gamma_2$, we can clearly see that $\varepsilon_B$'s spectrum is still symmetric but $\varepsilon_A$'s spectrum becomes asymmetric as shown in Fig.~\figDelta{e}. By setting $\Delta_k = 0$ and $\gamma_1 \neq \gamma_2$, $\varepsilon_A$'s spectrum becomes symmetric again as shown in Fig.~\figDelta{f}. Therefore, the asymmetry is caused by the combination of $\Delta_k \neq 0$ and $\gamma_1 \neq \gamma_2$.

\end{document}